**Novel dielectric anomalies due to spin-chains above and below Neél temperature in $Ca_3Co_2O_6$**

Tathamay Basu, Kartik K Iyer, Kiran Singh,[#] and E.V. Sampathkumaran

Tata Institute of Fundamental Research, Homi Bhabha Road, Colaba, Mumbai - 400005, India

[#]Current address: UGC-DAE Consortium for Scientific Research, University Campus, Khandwa Road, Indore - 452017, India

Correspondence and requests for materials should be addressed to E.V.S (sampath@mailhost.tifr.res.in).

**We bring out novel dielectric behavior of a spin-chain compound, $Ca_3Co_2O_6$, undergoing Neél order at $(T_N=)$ 24 K. It is found that the virgin curve in the plot of isothermal dielectric constant ($\varepsilon'$) versus magnetic-field lies outside the 'butterfly-shaped' envelope curve well below $T_N$ (e.g., 2.6 K), with a signature of a partial arrest of the high-field magnetoelectric (ME) phase in zero-field after travelling through magnetic-field-induced magnetic transitions. This behavior is in contrast to that observed in the isothermal magnetization data. Thus, this work brings out a novel case for 'phase-coexistence phenomenon' due to ME coupling. Another strange finding is that there is a weak, but a broad, peak in $\varepsilon'$ around 85 - 115 K well above $T_N$, attributable to incipient spin-chain magnetic ordering. This finding should inspire further work to study ME coupling on artificial assemblies of magnetic chains, also keeping in mind miniaturization required for possible applications.**

The search for materials with multiferroicity (for instance, those in which magnetic ordering and ferrolectricity are coupled) and with strong magnetolectric (ME) coupling has picked up momentum in recent years [1, 2] due to application potential of such materials as well as from the fundamental science angle. As a result of such a coupling between seemingly different phenomena, these two phases can intriguingly coexist under favorable circumstances [3], bearing relevance to the currently active topic of 'phase-coexistence phenomenon' in general in solids [4]. It is therefore of great interest to search for novel aspects of such a phase-coexistence situations arising out of this cross-coupling - an area which is yet to pick-up momentum. Another area remaining largely unexplored theoretically as well as experimentally is the investigation of ME coupling in *magnetically isolated* spin-chains, though this question was previously addressed in an organic one-dimensional quantum magnet and spin-Peierls instability was proposed to play a role on ME coupling [5]. Therefore, in order to enable magnetically controllable electric effects, for miniaturization of devices for applications commonly in particular, it is important to investigate whether the ME coupling within spin-chains (in the paramagnetic state) is in general more widespread without any restriction on spin values. Here, we demonstrate experimentally that the naturally occurring spin-chain oxide, $Ca_3Co_2O_6$ [6], with spin=2 on Co, turns out to be a novel material with 'exotic' dielectric properties, providing an ideal playground to probe both these aspects. We carried out complex dielectric measurements for this compound as a function of temperature (*T*) and magnetic-field (*H*). We find distinct intriguing features attributable to (i) a novel magnetoelectric phase-coexistence phenomenon following a magnetic-field-induced transition in the magnetically ordered state, and (ii) magnetoelectric coupling effect from spin-chains well above long range magnetic ordering.

The Co spin-chains are placed in a triangular fashion in this compound with a weak antiferromagnetic inter-chain interaction, naturally leading to geometrically frustrated magnetism at low temperatures. Geometrically frustrated magnetism and complex spin structures have been known to distort atomic arrangements thereby favouring inversion symmetry and hence ferroelectricity and/or ME coupling in the magnetically ordered state [see, for reviews, Refs. 7, 8]. In this respect, the compound $Ca_3Co_2O_6$ [9, 10] and some of its derivatives already attracted [10-13] some attention at the onset of three dimensional magnetic ordering. There are also some claims [14, 15] that this compound could be a promising cathode material for solid-state fuel cell applications as well as for various other applications such as pigments in coatings. Therefore, with geometrical frustration triggering a lot of investigations by several experimental methods as well as by theoretical studies [see, for instance, Refs. 16 - 47 and articles cited therein] due to novelties of the magnetic properties of this compound, this material clearly is of great interest from the fundamental physics as well as from the applications angle.

This compound crystallizing in a $K_4CdCl_6$-derived rhombohedral structure, consists (see figure 1) of chains of face-sharing $CoO_6$ trigonal prisms and $CoO_6$ octahedra running along c-axis with Ca separating the chains [6]. The former Co ions are in high-spin $d^6$ (*S*= 2) configuration, whereas the latter are in low-spin $d^6$ (*S*= 0) [16]. Broadly speaking, the three-dimensional magnetic ordering - the 'so-called' partially disordered antiferromagnetism (PDA) [17] - sets in at ($T_N$=) 24 K, with peculiar magnetic transitions near 10 and 7 K [18, 20, 30]. A noteworthy property [18, 20] is that there are equally-spaced (about 12 kOe) multiple jumps (rather sharp in single crystals) in isothermal magnetization, *M(H)*, at very low temperatures, mimicking quantum tunnelling behavior [19]. Dielectric anomalies have been reported in the virgin curve [9, 10, 30] at the magnetic transitions, induced by both *T* and *H,* which is usually taken [7, 48] as an evidence for magnetoelectric coupling effect and possible multiferroic behavior. Therefore, this system provides an ideal opportunity to address one of the issues addressed above - that is, to probe magnetoelectric phase coexistence phenomenon in zero-field after travelling through such magnetic-field-induced transitions. With respect to the second aim of this article, we would like to emphasize that this compound has been known to show interesting features, even well above $T_N$, in the range 80 - 100 K. For instance, [151]Eu Mössbauer spectrum (in the specimens doped with Eu) tends to undergo a gradual broadening [29], heat-capacity shows a broad bump [20], Extended X-ray Absorption Fine Structure (EXAFS) [36] and Raman [37] studies reveal structural anomalies etc. It turns out that the magnitude of the paramagnetic Curie temperature, which has a positive sign arising from intrachain ferromagnetic interaction, also falls in the same temperature range [18]. Hence, the magnetic anomalies around 80-100 K were interpreted in terms of 'incipient' spin-chain ordering, as a precursor to long-range magnetic order [29]. Therefore, the observed bond distortion anomalies also [36, 37] must be correlated to spin-chain ordering, which can also result in dielectric polarization effect. Since this compound is characterized by strong magnetoelectric effects below $T_N$, we considered it logical to choose this material to search for ME effect from incipient spin-chain ordering. The magnetic-moment containing chains can be treated 'isolated' magnetically, considering that intrachain magnetic coupling is much stronger than the strength of the interchain coupling [33] and the three-dimensional magnetic orderings sets in only at much lower temperatures.

## Results

Both the real part ($\varepsilon'$) and the loss part ($\tan\delta$) of the permittivity (Fig. 2a and Fig. 2b respectively for the curves at selected frequencies) show a huge frequency ($\nu$) dependence around 17-29K, typical of relaxor ferroelectrics [49]. There is a weak upturn beyond 100 K in both $\varepsilon'$ and $\tan\delta$, which is presumably due to a tendency to a small change in electrical resistivity. This compound has been known to be highly insulating, as indicated by very small values of $\tan\delta$ as well, and therefore no magnetoresistance and/or space charge contributions would contribute to dielectric constant [10], at least below 150 K. An inter-comparison of the data available in the literature [9, 10, 30] as well as the present data bring out that there are subtle differences in the shapes of curves $\varepsilon'(T)$. In fact, if the derivative of the curves at higher frequencies is viewed (see inset of Fig. 2), we observe distinct peaks, for instance, for $\nu$= 100 kHz, not only near 7 and 17 K as known earlier [9], but also near 25 K, which is very close to $T_N$. Following Kamiya and Batista [45], we believe that there are multiple features in the range 10 - 25 K due to the formation of microphases that could be sensitive to dielectric behavior. Clearly, the frequency dependencies of the resultant dielectric features are different and that is why one can not resolve it at all frequencies. It appears that the resolution of the features sensitively depends on sample e.g., with a possible role of grain boundaries. (We will address this issue more through Gd substitution studies as well in a future publication). The purpose of presenting above results is to stress that these peaks at the magnetic transitions in our data are well-separated from the high-temperature upturn and it is this resolution of the features in our specimen that enabled us to draw one of the main conclusions of this article.

We show in figure 3 the results of isothermal dielectric measurements performed with a frequency of 50 kHz, as a function of external magnetic-field at various temperatures (2.6, 5, 8 and 20 K) with the rate of change of *H* being 70 Oe/s. We measured the same with various frequencies in the range 10 - 50 kHz and all the features in general are found to be the same, though the absolute values vary with frequency (agreeing with the trends in figure 2). There is a small drop for an initial application of magnetic fields (<1 kOe) at very low temperatures (for instance, at 2.6 K), which may be due to possible grain boundary effects. For the sake of comparison, we plot in figures 3b and 3d experimentally measured *M(H)* loops at 2.6 K and 5 K in the same figure for the ZFC state (cooled from 100 K) of the specimen, with the same rate of change of *H*. The steps in *M*(*H*) mentioned at the introduction are not sharp (but appears as a change in slope only) in polycrystalline samples compared to the behavior in single crystals; corresponding anomalies $\varepsilon'(H)$ are rather weak and could be seen only in an expanded form of the virgin curve below 40 kOe in our sample (not shown in the figure in an expanded form, as this is not the subject of this article). For *T*= 8 and 20 K, such multiple steps in *M(H)* are absent and a single step at one-third of saturation magnetization, characteristic of such geometrically frustrated magnetic systems, has been known, followed by a sudden increase near 40 kOe, in the virgin curve. We see corresponding anomalies in $\varepsilon'(H)$ in figures 3c, 3e and 3f. The increase beyond 40 kOe is thus relatively more pronounced for all temperatures (see curve 1), compared to that at steps at lower fields. When the magnetic field is reversed from 140 kOe, there is a hysteresis for *T*= 2.6 - 8 K as a result of which the path-3 is followed and the zero-field value after this path is rather higher than that for the virgin state (Fig. 3a). If the magnetic-field direction is reversed and returned back to *H*= 0, paths 4 and 5 respectively are traced. Clearly, for the temperatures 2.6, 5 and 8 K, the virgin curves lie outside the envelope curve and butterfly-shaped loops are observed. The virgin curve can be restored, if the specimen is warmed up above $T_N$ and cooled again. We have also obtained the plots (see figures 3 h-j) after cooling the sample from 300 to 2.6 K in different magnetic fields and then switching off this field. It is noted that the curves corresponding to path-1 tend to move towards the envelope loop as the temperature value is increased. Thus, for instance, at 20 K, there is a sudden change in $\varepsilon'$ at the magnetic-field induced magnetic transition, but virgin state is restored after a reversal of the field is zero unlike the situation at further low temperatures.

We have made another intriguing observation in the data **above $T_N$**. Figure 4 is focussed on $\varepsilon'$ behaviour at various frequencies in the narrow range 50-130 K. We observe a well-defined peak, though broad and weak, in the range 85 - 115 K, and the intensity of this feature increases with ν. We think that its apparent absence at low frequencies is due to the fact that its presence is masked by the large change in the magnitude below 80 K and above 100 K (see figure 2). It is also interesting to note that the signal-to-noise ratio is further boosted with the application of an externally applied magnetic field, as shown in the same figure for 50 kOe, though absolute values are lowered in this temperature range, the reason for which is not clear to us at present.

We show in figure 5 the results of polarization studies. The curve below 50 K is shown in an expanded scale in the inset figure 5a. There are distinct features around 24, 10 and 5 K - that is, the temperatures near which magnetic transitions are observed - in the form of a peak, shoulder or a drop respectively. A fascinating finding is that there is a dramatic enhancement in the measured values of *P* with decreasing temperature near 90 K. This feature is highly reproduceable and the direction of polarization is found to be reversed after poling by negative electric fields. As stated earlier, electrical resistivity is sufficiently large that surface-charge related complications are not expected to contribute to this apparent transition. The field-dependent polarization (Fig 5, top inset) at 2.6 K also shows a change in slope around 40 -50 kOe, which can be directly correlated to metamagnetic transition in this compound.

**Discussion**

At the outset, it is worth recalling that this compound is characterized by a huge frequency dependence of the peak in *ac* susceptibility ($\chi_{ac}$) [17, 18, 19, 21], relaxation dynamics [19, 44] and memory effect [43], providing evidence for spin freezing phenomena below about 10 K. For instance [21], the peak in the real part of $\chi_{ac}$ shifts from ~10 K to ~17 K for a variation of ν from 1 Hz to 1 kHz in our polycrystalline sample. It should be stated that the large ν-dependence of the peak in $\varepsilon'$ (Figure 2) is similar to the behavior in *ac* $\chi_{ac}$. However, the lowest frequency used for the dielectric studies is the same as the highest one in *ac* $\chi$ [21]; for the lowest frequency employed here (1kHz), the peak temperature (17.5K) matches with that for 1 kHz in *ac* $\chi$. Thus, the peak-shift observed in $\varepsilon'(T)$ appears to be an extension of *ac* $\chi$ behavior and therefore the peak in $\varepsilon'(T)$, in our opinion, should be associated with the magnetic transition near 10-12 K. Considering huge ν-dependence in both these properties, we tend to believe that this compound could actually be classified as a 'multi-glass' - a phenomenon of great current interest [50-54].

We now discuss the two points of central emphasis:

In the plot of ε′(H) in figure 3, it was noted that the virgin curve lies outside the envelope loop, if the measurements are performed for the zero-field-cooled condition at very low temperatures. The butterfly-shaped curves with this behavior of virgin curve mimic those usually expected [55-58] in isothermal *M* and/or electrical resistivity after traversing through a (disorder-broadened) first-order field-induced magnetic transition and interpreted in terms of 'arrest' of the high-field phase when the magnetic-field is reduced to zero [56, 57]. Therefore, the observed shapes in figures 3a, 3c and 3e are signatures of (at least a 'partial') arrest of (high-field) magnetoelectric phase in zero field after travelling through the transitions around 40 kOe. We claim that the arrest is 'partial' because the zero-field value ultimately reached is less than the value at the highest-field. In other words, this zero-field state is a complex electric phase consisting of (magnetic) virgin component and high-field electric phase. At 20 K also, though there is a sudden change in ε′ at the magnetic-field induced magnetic transition, the virgin state is restored after a reversal of the field is zero, which implies that at this temperature, the 'magneto-electric phase co-existence phenomenon' appears to be negligible. In the plots shown in figures 3 h-j obtained after cooling the sample from 300 to 2.6 K in different magnetic fields, the signature of phase-coexistence diminishes after the magnetic-field is returned to zero. This establishes history-dependent behavior of the loop, similar to those traced in magnetization loops in other magnetic systems exhibiting first-order transitions [see, for instance, Refs. 55-58]. Thus, this work provides an experimental demonstration of magnetoelectric property for the proposal in Ref. 56 to observe phase-coexistence features for any phenomenon across first-order phase transitions. Another feature in the ε′(H) data in figure 3 is that the area under the envelope curve decreases from 2.6 to 8 K, which implies a reduction in the phase-coexistence region as a function of *T* and *H*. The observation we would like to stress and that was unexpected is that the *M(H)* plots, e.g., even at 2.6 and 5 K, do not show the signatures of 'arrest', as evidenced by the positioning of the virgin curves well-within the envelope curve, despite (disorder-broadened) first-order nature of the field-induced transitions. Though the exact origin of the difference between magnetoelectric and magnetization properties is not clear, it is possible that the apparent absence of this 'arrest' in magnetization data is an artifact of very complex relaxation phenomena and hierarchy of time scales of the low temperature state [see, for instance, Refs. 31, 32, 42], which could be different for magnetoelectric state and domain walls. Thus, the present results provide concrete evidence for a fascinating unforeseen situation in this compound in which a partial arrest of 'magnetoelectric phase' is unhindered, resulting in 'magnetoelectric phase-coexistence' without similar 'arrest' for the magnetic part. Various subtle details presented above reveal that there are further exotic opportunities to probe this phenomenon in depth, in particular, focussing on how the relaxation phenomena are different for such coexisting phases.

*With respect to the observation **above $T_N$***, that is, a well-defined broad peak in the range 85 - 115 K (figure 2), it is important to recall the information provided by other experimental methods in the literature. As stated in the introduction, EXAFS [36] and Raman [37] studies revealed structural distortions in the same *T*-range and other magnetic studies [20, 29] showed magnetic anomalies from spin-chains. Also, in this *T*-range, long-range magnetic ordering is absent and the intrachain ferromagnetic correlation strength is of similar magnitude as given by the magnitude of paramagnetic Curie temperature [18]. It is therefore quite natural to attribute the observed weak and broad dielectric anomaly to incipient spin-chain ordering, which appears to distort the distort the lattice by way of changing bond-distances (though no change of crystal symmetry has been reported in the literature). Thus, this work clearly demonstrates that ME effect can in principle arise from magnetically isolated chains for a spin value higher than 1/2 as well.

A fascinating finding is that there is a dramatic enhancement in the measured values of $P$ with decreasing temperature near 90 K, as though there is a ferroelectric transition induced by such an incipient spin-chain ordering. A compelling observation in favour of this inference is that the sign of $P$ was found to get reversed with a change of the direction of the poling electric-field and the magnitude also increases with electric-field. At present, we are not able to obtain loops of polarization versus electric-field to place this conclusion on firmer grounds. The above results are important, particularly noting that there is in general an interest to identify materials with magnetism-induced ferroelectricity above 77 K [59]. The polarization below 50 K (see the inset figure 5a) reveals distinct features at the magnetic transitions, presumably characterizing this compound as a multiferroic material. The field-dependent polarization (Fig 5, top inset) at 2.6 K shows a change in slope around 40 - 50 kOe, which can be directly correlated to metamagnetic transition in this compound. Thus these polarization results also bring out the existence of magnetoelectric coupling, both above and below $T_N$.

**Methods**

Polycrystalline specimen of $Ca_3Co_2O_6$ was prepared by solid state reaction method as described earlier [Ref. 21] starting from stoichiometric amounts of respective high purity (>99.99%) oxides, $CaCO_3$, and $Co_3O_4$. The sample was found to be single phase within the detection limit (<1%)) of x-ray diffraction ($CuK_\alpha$). The sample was further characterized for its homogeneity by Scanning Electron Microscope and Energy Dispersive X-ray Analysis as well as by magnetization studies. Complex dielectric permittivity measurements were carried out with several frequencies ($v$ = 1-100 kHz) at 1 V ac bias using E4980A LCR meter (Agilent Technologies) with a homemade sample-holder which is coupled to the Physical Properties Measurement System (PPMS, Quantum Design). Unless stated, all the measurements were performed during warming and for the zero-field-cooled (ZFC) condition of the specimen. The polarization ($P$) measurements were done employing an electrometer (Keithley, 6517A) with the same instrument as that for dielectric studies. For polarization measurements, the sample was cooled in an electric field of 400kV/m from 120 to 2K, and then the field was removed; subsequently, the sample was short-circuited to remove the surface charge (if any); we then waited for the decay of all external contribution and performed the charge measurement using Columbic mode of the electrometer as a function of temperature and magnetic-field.

*Note added: The experiments have been now performed by us on single crystals as well and the results obtained support the conclusions. Details of these results will be published elsewhere*

**Acknowledgements**

The authors would like to thank Professor S. Patnaik, Jawaharlal Nehru University, New Delhi, India, for verification of polarization data with a different set-up, and to Mr. Babu Gonde for his help in experimental part.

**Author contributions**

T.B. and K.K.I prepared the sample, characterized it, carried out experiments and analyzed the data. K.S. designed the sample-holder attachment to PPMS and standardized the set-up for dielectric studies. E.V.S. proposed the problem, formulated the paper and finalized in consultation with other authors.

**Additional information**

The authors declare no competing financial interests.

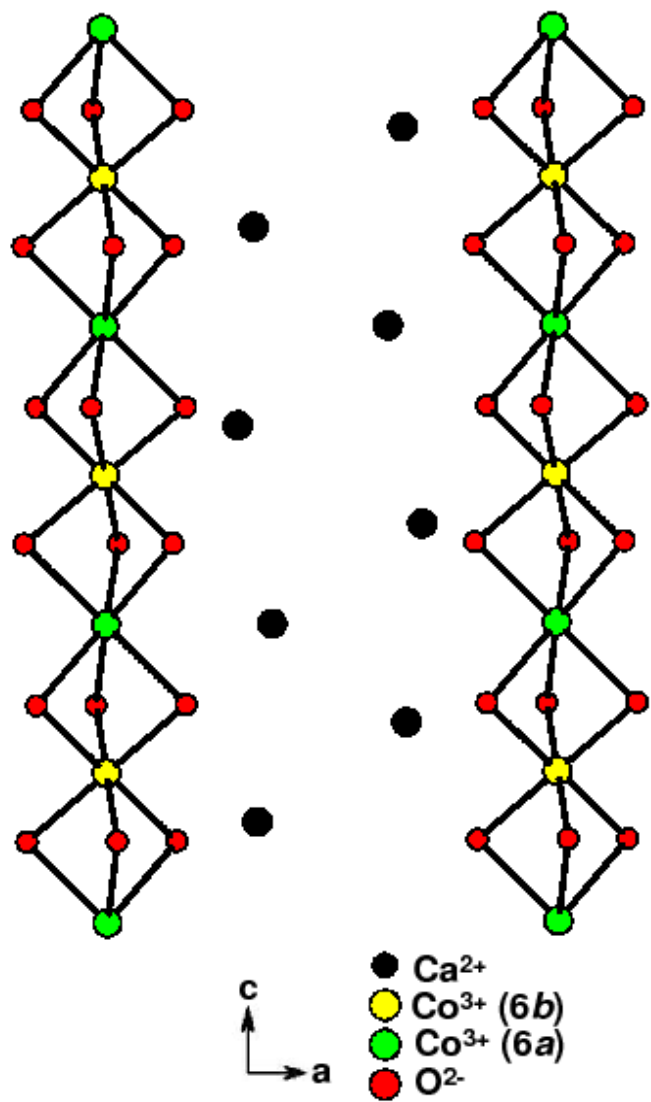

**Figure 1:**

The chain of Co ions viewed along *b*-direction in $Ca_3Co_2O_6$ is shown.

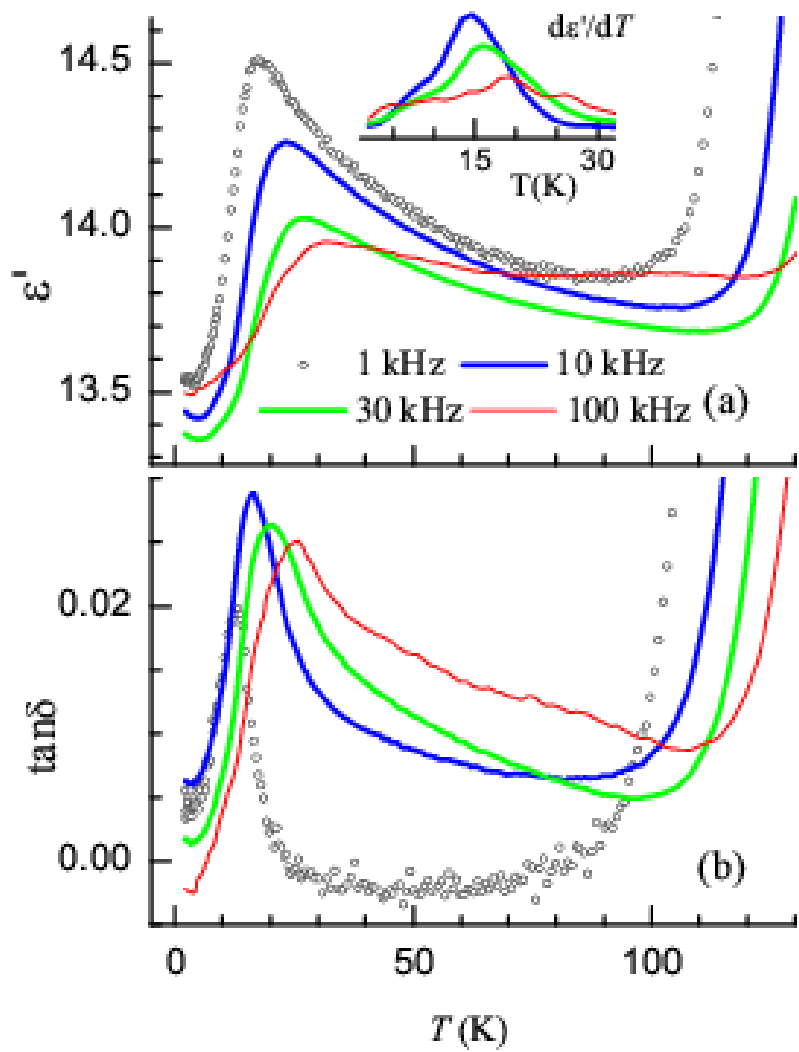

**Figure 2:**

(a) The real part of dielectric constant (ε′) and (b) loss factor (*tanδ*) as a function of temperature for $Ca_3Co_2O_6$. In the inset, the derivative plots are shown for some frequencies to highlight the contributions from additional transitions at higher frequencies.

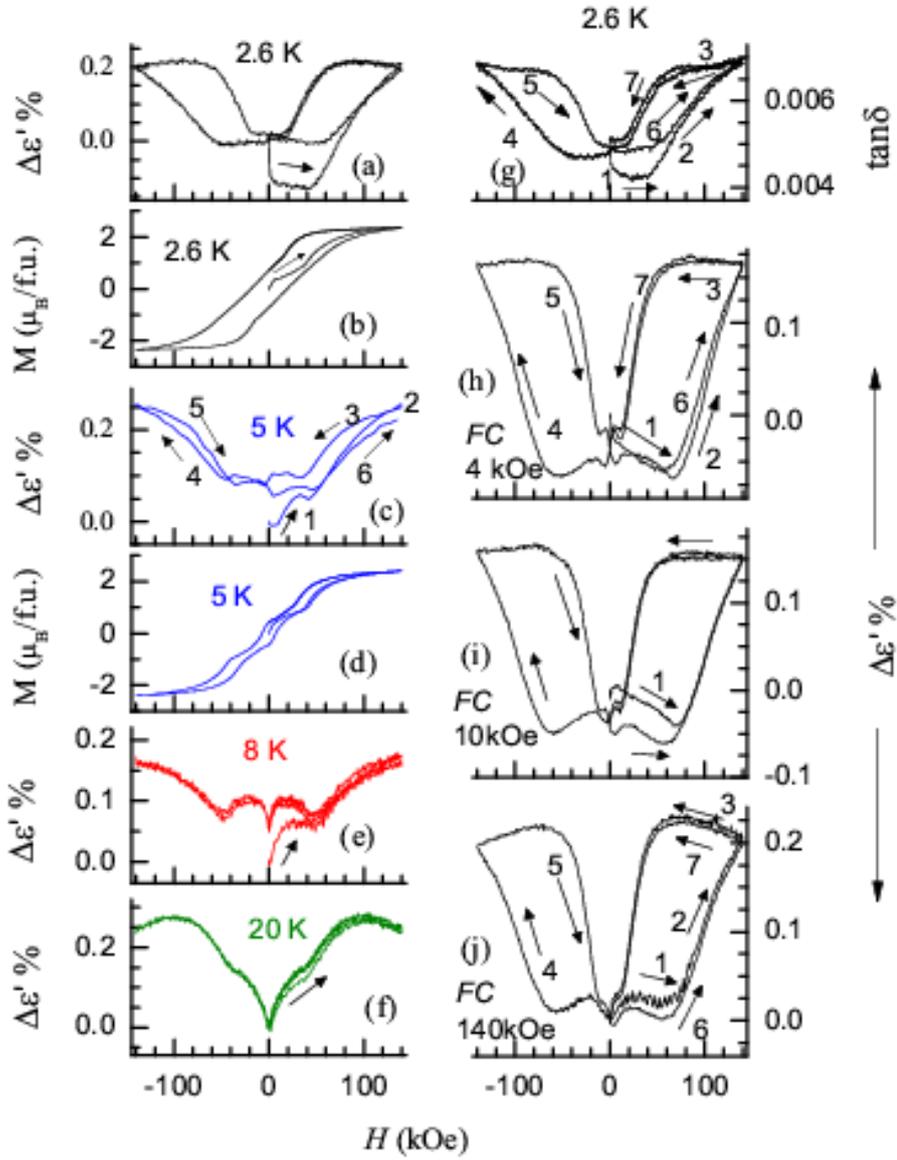

**Figure 3:** The change of dielectric constant [plotted in the form of $\Delta\varepsilon' = 100(\varepsilon'(H)-\varepsilon'(0))/\varepsilon'(0)$] as a function of magnetic field measured with a frequency of 50 kHz, for (a) 2.6 K, (c) 5 K, (e) 8 K, and (f) 20 K for $Ca_3Co_2O_6$ for the zero-field-cooled state of the specimen. Isothermal magnetization data are also plotted in (b) and (d). For 2.6 K, the zero-field *tanδ* curve is shown in (g) and the $\Delta\varepsilon'$ curves obtained for the specimen cooled in the presence of various fields are shown in (h), (i) and (j). The numericals and arrows serve as guides to the eyes to show the sequence in which the magnetic field was changed.

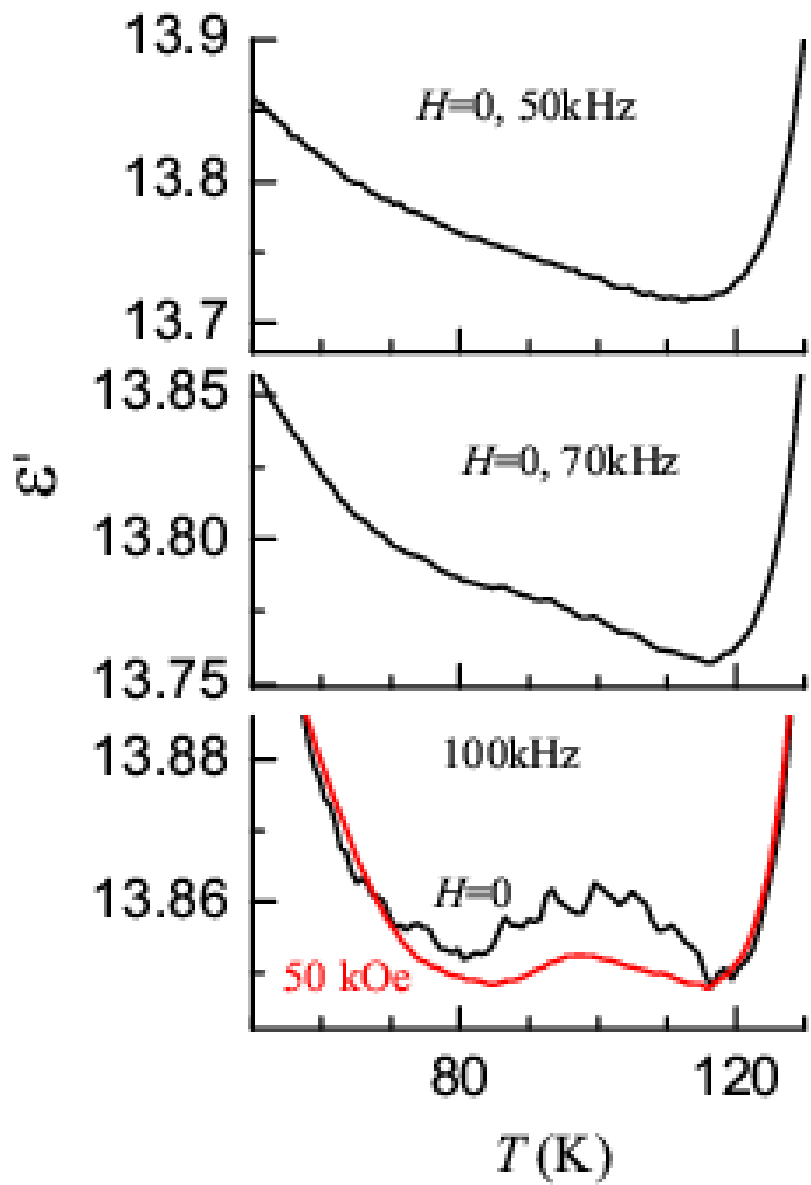

**Figure 4:** Dielectric constant as a function of temperature for $Ca_3Co_2O_6$ above 50 K for three frequencies in the absence of external magnetic field. The curve obtained in the presence of an external field of 50 kOe for 100 kHz is also shown.

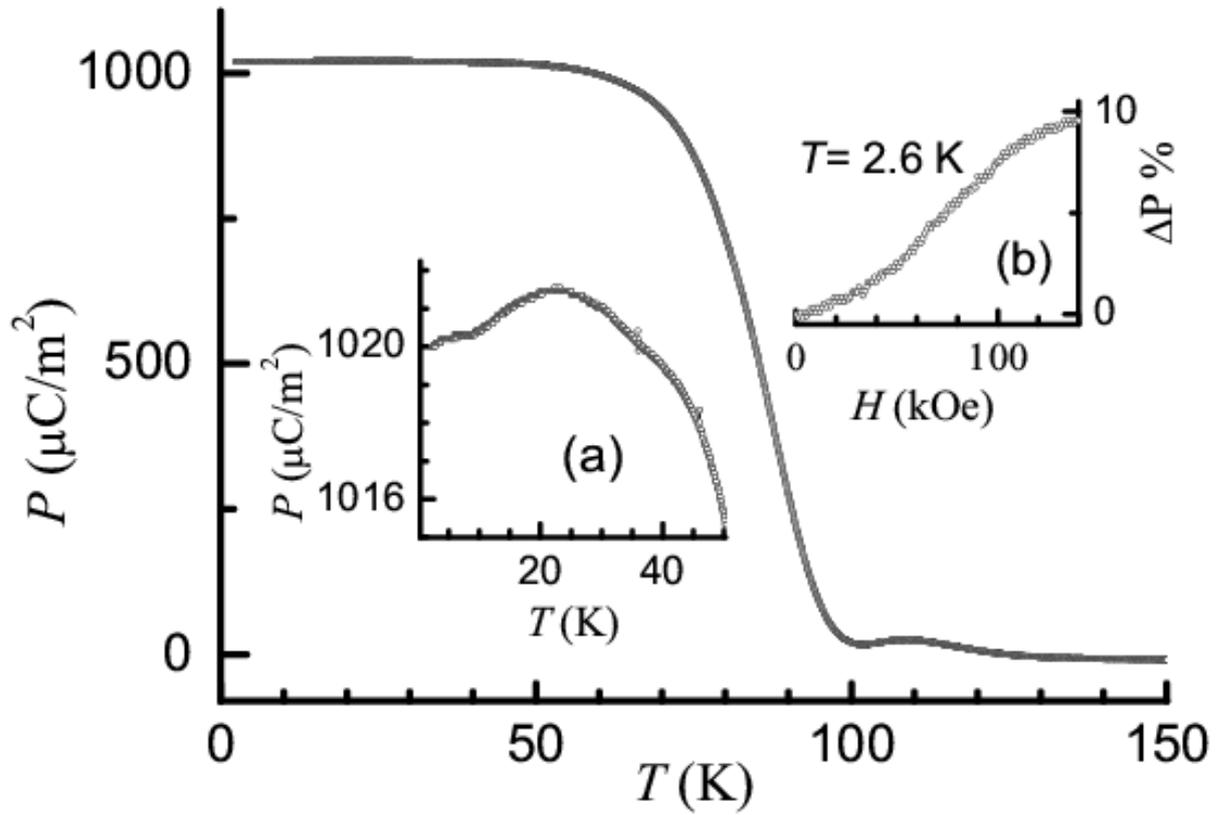

**Figure 5:** Polarization as a function of temperature (obtained as explained in the text) for $Ca_3Co_2O_6$. Inset (a) shows the low temperature data in an expanded form. The inset (b) shows polarization (plotted in the form of $\Delta P = [P(H)-P(0)/P(0)]$ as a function of magnetic field at 2.6 K.